# Combination of The Cellular Potts Model and Lattice Gas Cellular Automata For Simulating The Avascular Cancer Growth

Mehrdad Ghaemi[1], Amene Shahrokhi[2]

[1] Department of Chemistry, Teacher Training University, Tehran, Iran.
ghaemi@tmu.ac.ir
[2] District Health Center Shahre Ray, Vice-Chancellor for Health,Tehran University of Medical Science, Tehran, Iran.
amene_shahrokhi@hotmail.com

**Abstract.** The advantage of Cellular Potts Model (CPM) is due to its ability for introducing cell-cell interaction based on the well known statistical model i.e. the Potts model. On the other hand, Lattice gas Cellular Automata (LGCA) can simulate movement of cell in a simple and correct physical way. These characters of CPM and LGCA have been combined in a reaction-diffusion frame to simulate the dynamic of avascular cancer growth on a more physical basis.The cellular automaton is evolved on a square lattice on which in the diffusion step tumor cells (C) and necrotic cells (N) propagate in two dimensions and in the reaction step every cell can proliferate, be quiescent or die due to the apoptosis and the necrosis depending on its environment. The transition probabilities in the reaction step have been calculated by the Glauber algorithm and depend on the $K_{CC}$, $K_{NC}$, and $K_{NN}$ (cancer-cancer, necrotic-cancer, and necrotic-necrotic couplings respectively). It is shown the main feature of the cancer growth depends on the choice of magnitude of couplings and the advantage of this method compared to other methods is due to the fact that it needs only three parameters $K_{CC}$, $K_{NC}$ and $K_{NN}$ which are based on the well known physical ground i.e. the Potts model.

## 1 Introduction

Perhaps the most destructive phenomenon in natural science is the growth of cancer cells. The qualitative and quantitative comparison of simulated growth patterns with histological patterns of primary tumors may provide additional information about the morphology and the functional properties of cancer. Understanding the dynamics of cancer growth is one of the great challenges of modern science. The interest of the problem has led to the formulation of numerous growth models. Mathematical cancer modeling has been going on for many years. These models all included cancer cells and healthy cells to compete for space and nutrients, or drug. These progressed to Partial Differential Equation (PDE) models that generally modeled the tumor using diffusion of the cells [1]. Previous modeling techniques for the invasion process have included using sets of coupled reaction–diffusion equations for the cells and important groups of extracellular proteins and nutrients [2-5]. Today's model is typically a three dimensional PDE model with diffusion and advection for the cells, with scalar modifications based on nutrient and drug concentrations [6]. The PDE models can be numerically difficult to implement, however, due to a potentially high degree of coupling, besides the complex moving boundary problems. The inclusion of adhesion has been proven problematic in this type of model, although there have been some attempts [7, 8]. In addition, the reaction–diffusion approach makes the inclusion of the stochastic behavior of individual cells difficult to treat.

One way to circumvent this is to use a Cellular Automata (CA) model. CA approaches to biological complexity by describing specific biological models using two different types of cellular automata [9]: Lattice-Gas Cellular Automata (LGCA) and the Cellular Potts model (CPM).

LGCA can model a wide range of phenomena including the diffusion of fluids [10], reaction-diffusion processes [11], and population dynamics [12 ]. Dormann at al used LGCA for simulating dynamic of tumor growth [13]. In their model the dynamic of cancer growth can be explained as a reaction-diffusion process with three steps in each update. The reaction step contains mitosis, apoptosis, necrosis, and no change. In the diffusion step each cell moves to adjacent node according to it's velocity and in the redistribution step the occupation of channels in each site change according to preference weight. Dormann et al used phenomenological equations with adjustable parameters for the reaction part of the automata [14].

The Potts models [15] are general extension of the Ising model with $q$-state spin lattice, i.e., the Potts model with $q = 2$ reduces to Ising model. It attracted intense research interest in the 1970s and 1980s because it has a much richer phase structure and critical behavior than the Ising model [14]. In the cellular Potts model (CPM) [16-18] of cancer growth, each site contains one cell and considers necrotic, quiescent, and proliferating tumor cells as distinct cell types, in addition to healthy cells, with different growth rates and volume constraints for each type. In the CPM, transition probabilities between site states depend on both the energies of site-site adhesive and cell-specific non-local interactions.

The advantage of CPM is due to its ability for introducing cell-cell interaction in a correct and well known physical way. On the other hand LGCA can simulate movement of cell in a simple and correct physical way. In this article as explained in the next section, these characters of CPM and LGCA has been combined to simulate the dynamic of cancer growth on the more understandable and physical basis.

## 2 Method

The basic biological principles included in the model are cell proliferation, motility, necrosis, and apoptosis. The main body of the model is similar to the LGCA used for simulating reactive-diffusion systems. The cellular automaton evolves on a square lattice on which tumor cells (C) and necrotic cells (N) propagate in two dimensions. Each cell has associated with a velocity, which indicates the direction and the distance the cell will move in one time step. There are five velocity channels in each lattice site:

$V_0 = (0,0)$, $V_1 = (1,0)$, $V_2 = (0,1)$, $V_3 = (-1,0)$, $V_4 = (0,-1)$,

where $V_0$ is resting channel and $V_1$, $V_2$, $V_3$, and $V_4$ represent moving to right, up, left, and down, respectively. In each lattice site, we allow at most one cell (N or C) with each velocity, or maximum five cells in each lattice site. The dynamic is built from the following three basic steps: 1- the reaction step that consists of mitosis, apoptosis, necrosis, and no change, 2- the propagation step, and 3- the velocity redistribution step.

### 2-1 Reaction step

Every cell can proliferate, be quiescent, or die due to the apoptosis and the necrosis, depending on its environment. We have not enough and detailed information about the cell (cell itself is a complex system) and its interaction with other cells and materials, so deterministic prediction about the evolution of the cell is impossible and it is better to treat the cell dynamic as a stochastic dynamic. Cells adhere to each other by cell adhesion molecules (CAMs) which are present in the cell membrane. Usually cells of the same type have the same CAMs and adhere to each other more strongly than the cells of different types. Glazier and Graner [19] incorporated this type-dependent adhesion into the Potts model by assigning different coupling energies to different pairs of types.

Assume $C_{i,j}$ and $N_{i,j}$ are the number of cancer cells and necrotic cells in site $(i,j)$, respectively, and $K_{CC}$, $K_{NC}$ and $K_{NN}$ are cancer-cancer, necrotic-cancer and necrotic-necrotic couplings, respectively. For the sake of simplicity it is assumed that all cells in the same site interact with each others but there is no interaction between adjacent sites. Although it seems unrealistic but in the diffusion step the cells will move to the adjacent sites, and in the next time step each cell will interact with the cells which coming from the neighbours sites. So by evolving cellular automata each cell will experience the entire micro environment. The configuration energy of the lattice can be written as;

$$\frac{E_{conf}}{kT} = \sum_{i,j} E_{i,j,conf}$$

$$E_{i,j,conf} = -\left\{ \frac{1}{2} \left( C_{i,j}(C_{i,j}-1)K_{CC} + N_{i,j}(N_{i,j}-1)K_{NN} \right) + C_{i,j}N_{i,j}K_{NC} \right\}, \quad (2)$$

where $E_{i,j,conf}$ is the configuration energy of the site $i, j$ and $k$ is the Boltzmann constant. Cell-cell interactions are adhesive, thus the couplings are positive (note that there is a minus sign before bracket in the Eq. 2). Now in each lattice site one of the following reactions can occur at each time step;

Quiescent :

$$\begin{cases} C_{i,j} \longrightarrow C_{i,j} \\ N_{i,j} \longrightarrow N_{i,j} \end{cases}$$

Proliferation:

$$\begin{cases} C_{i,j} \longrightarrow C_{i,j}+1 \\ N_{i,j} \longrightarrow N_{i,j} \end{cases} \text{ if and only if } (C_{i,j}+N_{i,j}<5 \text{ and } C_{i,j} \geq 1)$$

Apoptosis:

$$\begin{cases} C_{i,j} \longrightarrow C_{i,j}-1 \\ N_{i,j} \longrightarrow N_{i,j} \end{cases} \text{ if and only if } (C_{i,j} \geq 1)$$

Necrosis:

$$\begin{cases} C_{i,j} \longrightarrow C_{i,j}-1 \\ N_{i,j} \longrightarrow N_{i,j}+1 \end{cases} \text{ if and only if } (C_{i,j} \geq 1)$$

By replacing the right hand side variables of each reaction with the previous one in eq. 2 we can compute the corresponding configuration energy and by method use the Glauber algorithm [20] the probability of each reaction in the each lattice site can be computed. For example

$$P_{apoptosis} = \frac{e^{\frac{-E_{apoptosis}}{kT}}}{e^{\frac{-E_{quiescent}}{kT}} + e^{\frac{-E_{prolifrations}}{kT}} + e^{\frac{-E_{apoptosis}}{kT}} + e^{\frac{-E_{necrosiss}}{kT}}}, \quad (3)$$

where each term in the right hand side of eq. 3 is a Boltzmann factor. According to restriction of maximum five cells in each lattice site and non negative values of $C_{i,j}$ and $N_{i,j}$, in some cases one or more of the reactions cannot be occur. For these cases we set the corresponding Boltzmann factor equal to zero.

**2-2 Propagation and redistribution steps**

In the propagation step each cell will move to neighbor site according to its velocity. Because the cells collide with each other the velocity of the cell should be changed. In addition, according to the chemotaxic effect, the cancerous cell will move toward the source of the chemotaxic materials i. e. the necrotic cells. We can include these effects in the redistribution step. In this step the velocity of the cancerous cells and the necrotic cells are changed according to the following rules:

a) Because the necrotic cells are less motile compared to the cancerous cells, first the velocity of the necrotic cells is redistributed then the cancerous cells are redistributed over the remainder channels.
b) Due to the adhesion effect the resting channel ($V_5$) is filled first and the remainder cells are distributed among the channels $V_1$ to $V_4$ according to the probability of occupation of channels. This probability is proportional to the gradient of the concentration of the chemotaxic materials. So in the simplest case we can assume that the relative magnitude of these probabilities is equal to the relative number of the necrotic cells in the adjacent sites:

$$\frac{P_i}{P_j} = \frac{n_i}{n_j}, \quad (4)$$

where $P_i$ is the probability of occupation of the channel $V_i$, and $n_i$ is the number of necrotic cells in the adjacent site conjugate to channel $V_i$.

## 3 Results and Discussion

The simulation is conducted on a 600 × 600 square lattice with central site initially defined to contain five cancerous cells. The size of the lattice is chosen sufficiently large such that the boundaries do not influence the tumor growth within the considered time interval. Multicellular spheroids have a well-established characteristic structure. There is an outer rim of proliferating cells (a few hundred $\mu$m thick) and an inner core of necrotic cells. Between these there is a layer of quiescent cells, which are not dividing but are alive, and can begin dividing again if environmental conditions change. The choice of coupling parameters values ($K_{CC}$ = 3, $K_{NC}$ = 1.5 and $K_{NN}$ = 3 ) are determined in such a way to produce multicellular spheroids shape (Fig. 1). The results show that by increasing the value of $K_{NC}$, the diameter of the layer of quiescent cells will decrease more rapidly and simultaneously the rate of growing of the inner core of necrotic cells will increase (Fig. 2). The future of tumor strongly depend on the values of $K_{CC}$ and $K_{NN}$. For the values of $K_{CC}$ = $K_{NN}$ < 2.5 the tumor initially grow up and after some time step the layers of proliferating and quiescent cell will be destroyed (Fig. 3).

The average number of cancerous cell versus time step is calculated for 20 different samples with the coupling parameters $K_{CC}$ = 3, $K_{NC}$ = 1.5 and $K_{NN}$ = 3 (Fig. 4). After an initial exponential growth phase, growth significantly slows down.

As it is seen the main feature of the cancer growth can be obtained by the combination of CPM and LGCA. The advantage of this method compared to other simulation of cancer growth is that the present method needs only three parameters $K_{CC}$, $K_{NC}$ and $K_{NN}$ based on the well known physical ground i.e. the Potts model. This simulation has been greatly simplified by neglecting some effects such as: interaction of healthy cells with cancerous cells, the effect of nutrients concentrations and limited volume space for tumor and it seems the addition of these effects is not problematic in this simulation.

## Acknowledgment

we acknowledge Prof. R. Islampour for his useful comments.

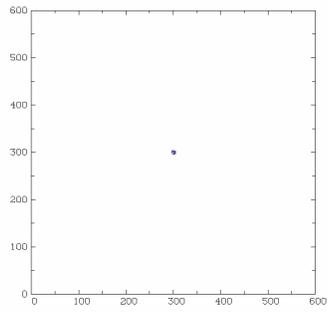
After 5 time steps

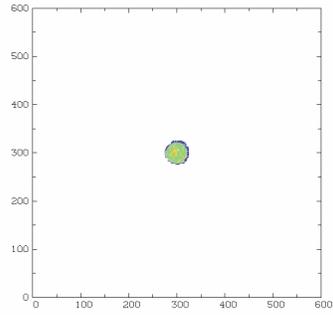
After 50 time steps

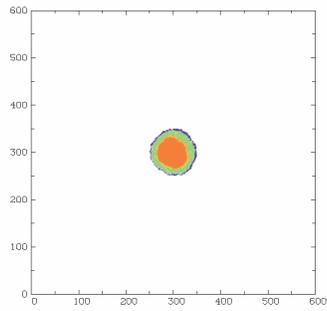
After 100 time steps

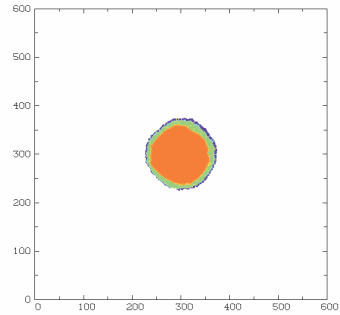
After 150 time steps

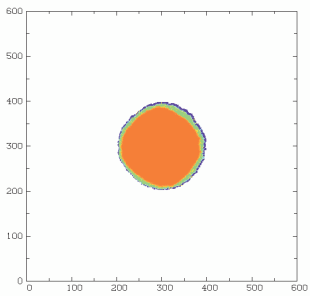
After 200 time steps

Figure 1- The pattern of cancer growth on the 600 × 600 square lattice using coupling parameters $K_{CC} = 3$, $K_{NC} = 1.5$ and $K_{NN} = 3$ at different time steps. Red, green and blue colors correspond to necrotic, quiescent and proliferating shells respectively.

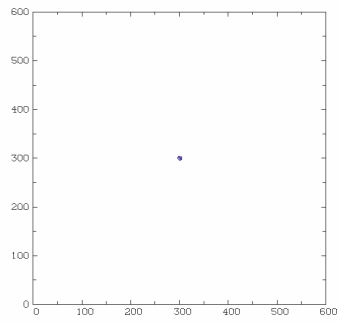
After 5 time steps

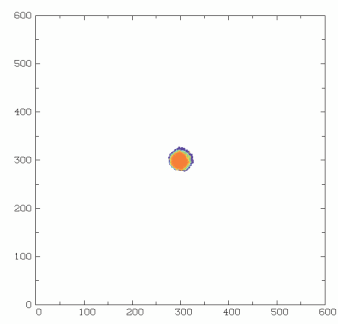
After 50 time steps

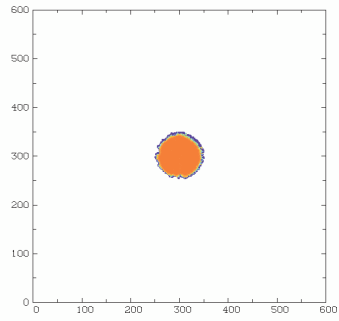
After 100 time steps

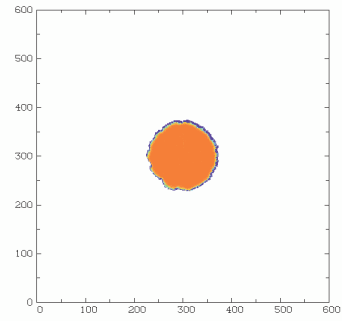
After 150 time steps

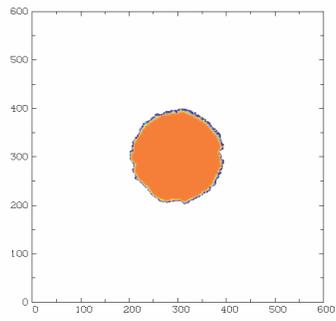
After 200 time steps

Figure 2- Same as Figure 1 but with $K_{CC} = 3$, $K_{NC} = 2.5$ and $K_{NN} = 3$.

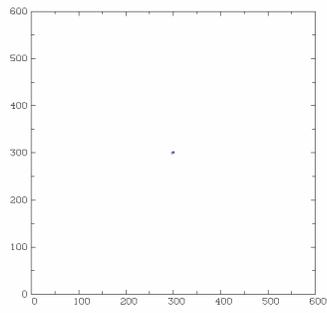
After 5 time steps

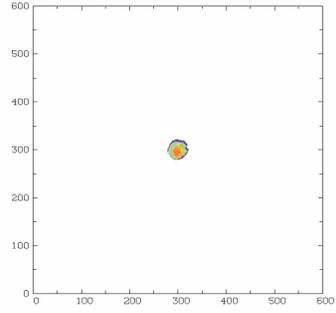
After 50 time steps

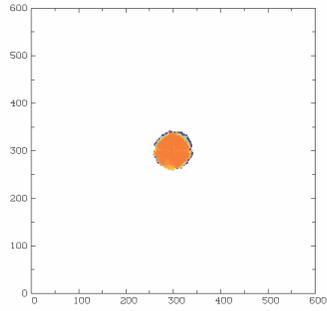
After 100 time steps

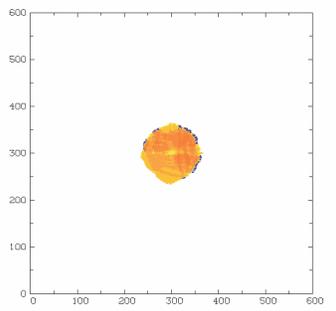
After 150 time steps

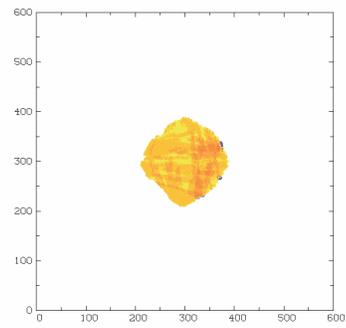
After 200 time steps

Figure 3- Same as Figure 1 but with $K_{CC} = 2$, $K_{NC} = 0$ and $K_{NN} = 2$.

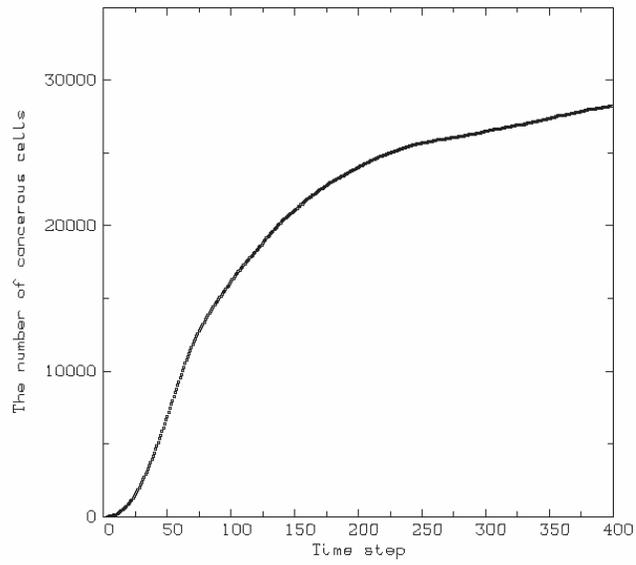

Figure 4 – The average number of cancerous cell versus time step for 20 different samples using coupling parameters $K_{CC} = 3$, $K_{NC} = 1.5$ and $K_{NN} = 3$.